\documentclass[a4paper,fleqn,usenatbib,useAMS]{mnras}

\usepackage[T1]{fontenc}
\usepackage{ae,aecompl}     
\usepackage{graphicx}
\newcommand{\teff}{${T}_{\mathrm{eff}}$}
\newcommand{\logg}{$\log{g}$}
\newcommand{\msun}{$M_{\odot}$}

\def\aap{A\&A}
\def\apjl{ApJ}
\def\apj{ApJ}
\def\apjs{ApJS}
\def\aj{AJ}
\def\mnras{MNRAS}

\def\pasp{PASP}

\def\araa{ARA\&A}

\renewcommand{\arraystretch}{1.1}
\title[ZZ Cetis in the original Kepler mission]{The search for ZZ Ceti stars in the original {\em Kepler} mission}

\author[Greiss et al.]{S.~Greiss,$^{1}$
J.~J.~Hermes,$^{1,2,3}$\thanks{E-mail: jjhermes@unc.edu}
B.~T.~G\"{a}nsicke,$^{1}$
D.~T.~H.~Steeghs,$^{1}$
Keaton~J.~Bell,$^{4}$
\newauthor
R.~Raddi,$^{1}$
P.-E.~Tremblay,$^{1}$
E.~Breedt,$^{1}$
G.~Ramsay,$^{5}$
D.~Koester,$^{6}$
P.~J.~Carter,$^{1,7}$
\newauthor
Z.~Vanderbosch,$^{4}$
D.~E.~Winget,$^{4}$ and
K.~I.~Winget$^{4}$
\\
$^{1}$Department of Physics, University of Warwick, Gibbet Hill Road, Coventry, CV4 7AL, UK\\
$^{2}$Department of Physics and Astronomy, University of North Carolina, Chapel Hill, NC\,-\,27599-3255, USA \\
$^{3}$Hubble Fellow \\
$^{4}$Department of Astronomy, University of Texas at Austin, Austin, TX - 78712, USA \\
$^{5}$Armagh Observatory, College Hill, Armagh, BT61 9DG\\
$^{6}$Institut f{\"u}r Theoretische Physik und Astrophysik, University of Kiel, 24098 Kiel, Germany \\
$^{7}$School of Physics, H.H.\ Wills Physics Laboratory,University of Bristol, Tyndall Avenue, Bristol BS8~1TL}

\begin{document}

\maketitle

\label{firstpage}

\begin{abstract}
We report the discovery of 42 white dwarfs in the original {\em Kepler} mission field, including nine new confirmed pulsating hydrogen-atmosphere white dwarfs (ZZ\,Ceti stars). Guided by the {\it Kepler}-INT Survey (KIS), we selected white dwarf candidates on the basis of their $U-g$, $g-r$, and $r-H\alpha$ photometric colours. We followed up these candidates with high-signal-to-noise optical spectroscopy from the 4.2-m William Herschel Telescope. Using ground-based, time-series photometry, we put our sample of new spectroscopically characterized white dwarfs in the context of the empirical ZZ\,Ceti instability strip. Prior to our search, only two pulsating white dwarfs had been observed by {\em Kepler}. Ultimately, four of our new ZZ\,Cetis were observed from space. These rich datasets are helping initiate a rapid advancement in the asteroseismic investigation of pulsating white dwarfs, which continues with the extended {\em Kepler} mission, {\em K2}.
\end{abstract}

\begin{keywords}
asteroseismology, surveys, stars: white dwarfs, oscillations
\end{keywords}

\section{Introduction}

White dwarfs are the end points of all low- to intermediate-mass stars, which are the majority of stars in the Universe. They are dense stellar remnants composed of electron degenerate cores surrounded by non-degenerate envelopes. These evolved stars provide key insight into Galactic star formation and evolution.

To use white dwarfs as tracers of Galactic evolution, we must determine their basic physical parameters, such as effective temperatures ($T_\mathrm{eff}$), surface gravities ($\log\,g$) and masses. More than 80\,per\,cent of white dwarfs have hydrogen-rich atmospheres, known as DA white dwarfs \citep{Giammichele12,Kleinman13}. Spectroscopy has been a successful tool in obtaining DA atmospheric parameters, yet it only provides a view of the outermost layers of the white dwarf, and it is subject to systematic problems that require correction if the star is cooler than roughly $13\,000$\,K and the surface is convective \citep{Tremblay10,Tremblay13}.

Asteroseismology, however, can probe deep into the interior of a white dwarf and provide information on its composition, rotation period, magnetic field strength, mass, temperature and luminosity, by matching the observed non-radial $g$-mode pulsations to theoretical models (see reviews by \citealt{WinKep08, FontBrass08, Althaus10}).

As white dwarfs cool, they pass through instability strips depending on their outermost envelope composition, which coincides with the onset of a partial ionization zone. This zone efficiently drives pulsations, which cause periodic brightness variations of the white dwarf \citep{Brickhill91}. The variable DA white dwarfs (DAVs), also known as ZZ\,Ceti stars, are the most commonly found and studied type of pulsating white dwarfs. Their effective temperatures range between $\simeq12\,300-10\,900$\,K for a typical mass of M $\simeq$ 0.6\,\msun, with pulsation periods ranging from $100-1400$\,s \citep{Mukadam06,Gianninas11}.

Ground-based studies of ZZ\,Ceti stars have been carried out for decades, but very few have been observed long enough to resolve more than six pulsation modes. Crucially, there are potentially nine free parameters in the asteroseismic modeling of pulsating white dwarfs \citep{Bradley01}, which has required holding fixed many parameters in order to constrain the internal properties of ZZ\,Cetis. It is clear that fully constraining these free parameters and thus the internal white dwarf structure and evolution will require a larger sample of rich asteroseismic observations.

Considerable effort has been expended to obtain uninterrupted photometry from coordinated, multi-site, ground-based campaigns to study pulsating white dwarfs, especially through the Whole Earth Telescope \citep{Nather90}, which has been operating for more than two decades (e.g., \citealt{Winget91,Provencal12}). However, the {\em Kepler} planet-hunting spacecraft offers a unique opportunity to obtain high-quality, space-based light curves of variable stars, white dwarfs included.

Unfortunately, few white dwarfs were known in the original {\em Kepler} mission field, and only two pulsating white dwarfs were discovered within the first two years of the mission \citep{Ostensen11a, Hermes11}. In order to increase this sample, we began the search for all white dwarfs in the original field of the {\em Kepler} mission, and more specifically any possible ZZ\,Ceti stars. We created the {\it Kepler}-INT Survey (KIS, \citealt{Greiss12a}) in order to select white dwarf candidates using colour-colour diagrams, and report on that search here.

In Sections \ref{selection} and \ref{spec}, we describe the selection method of our white dwarfs and present their spectroscopic observations. In Section \ref{astero}, we focus on the nine new pulsating white dwarfs. We conclude in Section \ref{conclusion}.

\section{Target selection}
\label{selection}

\begin{figure}
\noindent
\parbox{\columnwidth}{
\centering
\hspace*{-0.5cm}
\includegraphics[angle=0, trim=0cm 0cm 0cm 0cm, clip, scale=0.8]{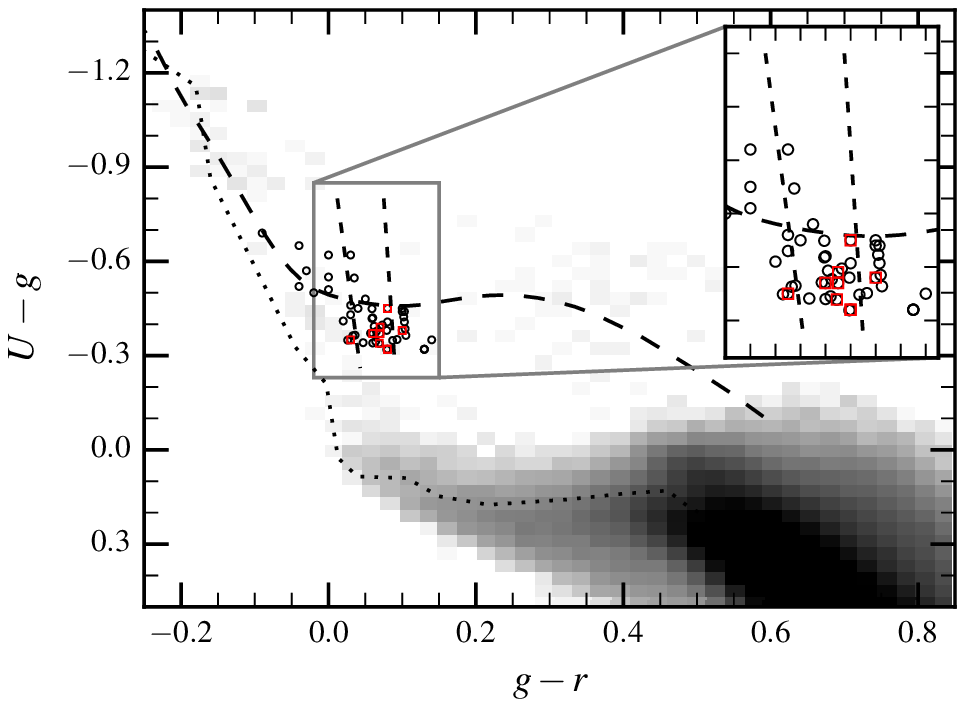}}
\\
\smallskip
\parbox{\columnwidth}{
\centering
\vspace*{-0.2cm}
\hspace*{-0.5cm}
\includegraphics[angle=0, trim=0cm 0cm 0cm 0cm, clip, scale=0.8]{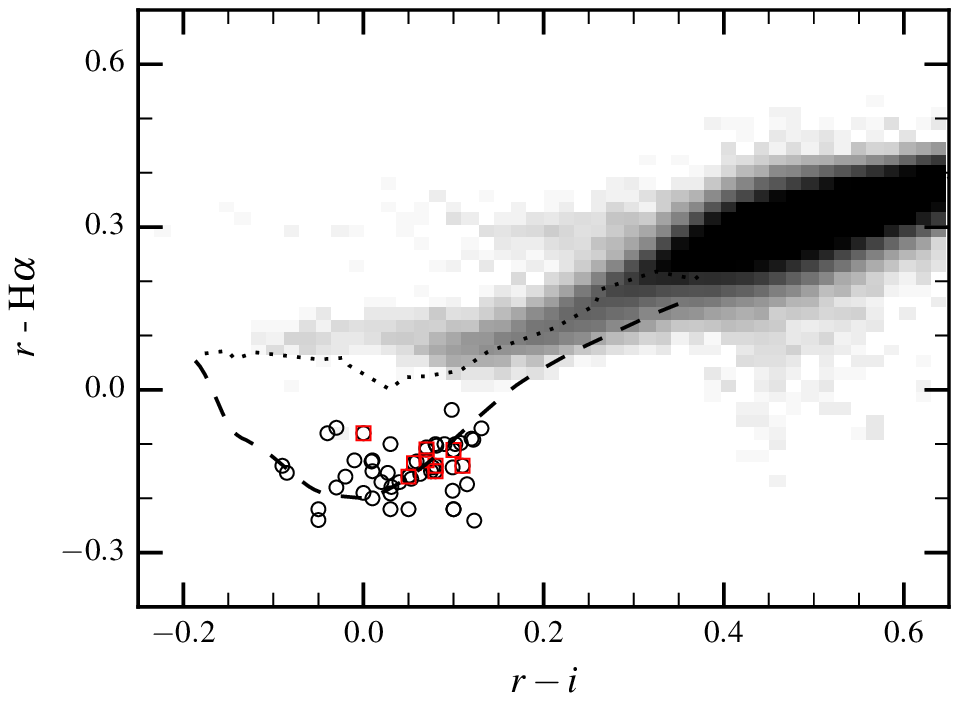}}
\vspace*{-0.4cm}
\caption{\small $U-g$, $g-r$ (top) and $r-{\mathrm H}\alpha$, $r-i$ (bottom) colour-colour diagrams of point sources from the \textit{Kepler}-INT survey (gray scale), with \logg\ $= 8.0$ DA white dwarf cooling tracks overplotted as a dashed line (the dotted line marks the region expected of main-sequence dwarfs). Narrowing our colour selection around the first ZZ\,Ceti in the field identified by \citet{Hermes11} left $\simeq$60 white dwarf candidates (open circles). The red open squares correspond to the nine ZZ\,Ceti stars we confirmed in the {\it Kepler} field; two have identical $U-g$, $g-r$ colours.\label{fig:colours}}
\end{figure}

We selected our white dwarf candidates using $U-g$, $g-r$ and $r-{\mathrm H}\alpha$, $r-i$ colour-colour diagrams using data from the {\it Kepler}-INT Survey (KIS), shown in Fig.~\ref{fig:colours}. KIS is a deep optical survey using the Wide Field Camera on the 2.5-m Isaac Newton Telescope (INT), taken through four broadband filters ($U, g, r, i$) and one narrowband filter (H$\alpha$), covering more than 97\,per\,cent of the original {\em Kepler} field down to $\sim$20$^{th}$ mag \citep{Greiss12a,Greiss12b}. All magnitudes for the KIS survey are expressed in the Vega system.

White dwarfs have bluer colours than main-sequence stars, and most single DA white dwarfs also have strong H$\alpha$ absorption lines, leading to $r-{\mathrm H}\alpha < 0$ (see bottom panel of Fig.~\ref{fig:colours}). We have integrated the atmospheric models of canonical-mass 0.6\,\msun\ (\logg\ $=8.0$) white dwarfs of \citet{Koester10} with the various filter profiles to guide our colour cuts, similar to \citet{Groot09}.

Our photometric selection recovered KIC\,4552982, the first ZZ\,Ceti star in the {\it Kepler} field \citep{Hermes11,Bell15}. We narrowed our selection to a small region around KIC\,4552982 and to candidates close to the empirical ($T_\mathrm{eff}$, $\log$~$g$) instability strip projected into $U-g, g-r$ space. This left more than 60 white dwarf candidates, 43 of which we were able to follow up spectroscopically (Table~\ref{phot-summary}).

\section{Spectroscopy}
\label{spec}

\subsection{WHT Observations}

We were awarded a total of eight nights on the 4.2-m William Herschel Telescope (WHT) in 2012, 2013, and 2014, where we obtained intermediate resolution spectra of 43 of our candidates in order to confirm their identities as white dwarf stars and to characterize their atmospheric parameters, especially their effective temperatures and surface gravities. We used the Intermediate-dispersion Spectrograph and Imaging System\footnote{http://www.ing.iac.es/Astronomy/instruments/isis/} (ISIS), with the R600R and R600B gratings on the red and blue arms, respectively.

The slit widths were chosen close to the seeing of each night to maximize spectral resolution ($\simeq1.5-2.0$\,\AA). A full journal of observations is included in Table~\ref{tab:specjour}. The blue arm was centred at 4351\,\AA\ and the red arm at 6562\,\AA. The spectra covered a wavelength range from roughly $3800-5100$\,\AA\ in the blue (see Fig.~\ref{spec-fit}), and roughly $5600-7100$\,\AA\ in the red. However, since the higher-order Balmer lines contain the most information about the atmospheric parameters (e.g., \citealt{Bergeron92}), we only used the blue arm for our Balmer profile fits (Section~\ref{sec:atmosparams}).

% +++++++++++++++++++++++++++ Phot Journal +++++++++++++++++++++++++++ %
\begin{table}
\begin{center}
 \centering
  \caption{Journal of spectroscopic observations.}\label{tab:specjour}
  \begin{tabular}{@{}lcccc@{}}
  \hline
Night & UT Date & Slit & Seeing        & Notes \\
      &         & (\arcsec) & (\arcsec) &  \\
 \hline
a & 2012 August 9  & 1.2 & 1.6          & Clouds \\
b & 2012 August 10 & 1.2 & $2.5-4.5$ & Haze \\
c & 2012 August 11 & 1.2 & $0.7-1.1$ & Thin \\
d & 2013 June 6    & 1.0 & $0.8-1.0$ & Clear \\
e & 2013 June 7    & 1.0 & $0.8-1.0$ & Clear \\
f & 2013 June 8    & 1.0 & $0.8-1.0$ & Clear \\
g & 2014 July 25   & 0.8 & $0.6-0.8$ & Clear \\
h & 2014 July 26   & 0.8 & $0.4-0.6$ & Clear \\

\hline
\end{tabular}
\end{center}
\end{table}
% ==================================================================== %

All our spectra were de-biased and flat-fielded within standard STARLINK\footnote{The STARLINK Software Group homepage website is http://starlink.jach.hawaii.edu/starlink} packages KAPPA, FIGARO and CONVERT. They were then optimally extracted using PAMELA\footnote{PAMELA was written by T. R. Marsh and can be found in the STARLINK distribution Hawaiki and later releases.} \citep{Marsh89}. Copper-argon arc lamp exposures were taken at the start and end of each night for the wavelength calibration of the spectra. We also observed several standard stars at the beginning, middle and end of each night: Feige 34, Grw+70 5824, LB 1240, G191-B2B and L1512-34. We used MOLLY\footnote{MOLLY was written by T. R. Marsh and is available from http://www.warwick.ac.uk/go/trmarsh/software.} for the wavelength and flux calibration of the extracted 1-D spectra obtained. 

All 43 observed targets were confirmed to be DA white dwarfs, verifying our colour-selection methods. We provide the coordinates and $U,g,r,i,H\alpha$ magnitudes of our targets in Table~\ref{phot-summary}, and detail the spectroscopic observations and the derived atmospheric parameters in Table~\ref{spec-summary}. Only one object we observed was a previously known white dwarf: KISJ1909+4717, also known as KIC\,10198116, which was previously discovered by \citet{Ostensen11b} and observed for one month with {\em Kepler}.

\subsection{Atmospheric Parameters}
\label{sec:atmosparams}

We were primarily interested in the potential ZZ\,Ceti stars within our DA white dwarf sample, and spectroscopically determining temperatures is the most efficient way to find new pulsating white dwarfs (e.g., \citealt{Mukadam04}). Because pulsations in ZZ\,Cetis are driven by a hydrogen partial-ionization zone they are confined to a narrow instability strip in (\teff,\logg) space \citep{Gianninas11}, making it possible to select ZZ\,Ceti candidates from our sample of white dwarfs on the basis of those parameters. 

\begin{table*}
\caption{Summary of photometry of our candidates selected for spectroscopic follow-up. All magnitudes are in the Vega system. \label{phot-summary}}
\scriptsize
\begin{tabular}{lrrrrrrrrr}
\hline
KIS ID     &  KIC ID   &  RA (J2000)  &  Dec (J2000) & $U$ (mag) & $g$ (mag) & $r$ (mag) & H$\alpha$ (mag) & $i$ (mag)  & Comments \\
\hline
J1846+4157 &  -        &  18 46 35.96 &  +41 57 07.6 & 17.96(2) & 18.31(1) & 18.22(1) & 18.33(3) & 18.14(2) &  - \\ 
J1848+4225 &  6923777  &  18 48 30.83 &  +42 25 15.5 & 18.02(1) & 18.36(1) & 18.32(1) & 18.51(4) & 18.29(2) &  NOV $>$3.5\,ppt \\ 
J1851+4506 &  -        &  18 51 01.28 &  +45 06 49.5 & 18.51(4) & 18.92(3) & 18.81(3) & 18.96(9) & 18.72(4) &  - \\ 
J1857+4908 &  11337510 &  18 57 30.66 &  +49 08 36.2 & 18.34(1) & 18.79(1) & 18.75(1) & 18.90(3) & 18.74(2) &  - \\ 
J1858+4613 &  9573820  &  18 58 10.23 &  +46 13 07.9 & 18.31(2) & 18.77(1) & 18.74(2) & 18.98(5) & 18.79(4) &  - \\ 
J1859+4842 &  11125021 &  18 59 01.98 &  +48 42 38.3 & 18.57(1) & 19.02(1) & 18.92(1) & 19.09(2) & 18.88(1) &  NOV $>$6.3\,ppt \\ 
J1902+4223 &  -        &  19 02 07.12 &  +42 23 25.5 & 18.67(1) & 18.99(1) & 18.86(2) & 19.08(5) & 18.76(3) &  - \\ 
J1904+4130 &  -        &  19 04 50.16 &  +41 30 16.7 & 17.03(2) & 17.45(1) & 17.39(1) & 17.54(3) & 17.36(1) &  - \\ 
J1904+4245 &  7184288  &  19 04 26.62 &  +42 45 48.7 & 18.21(1) & 18.78(1) & 18.81(1) & 18.97(4) & 18.83(3) &  - \\ 
J1906+4354 &  8084967  &  19 06 31.31 &  +43 54 48.8 & 18.67(2) & 19.01(1) & 18.95(2) & 19.17(6) & 19.00(4) &  - \\ 
J1906+5002 &  11805054 &  19 06 34.71 &  +50 02 17.0 & 18.73(3) & 19.28(2) & 19.24(3) & 19.39(8) & 19.33(5) &  - \\ 
J1908+4316 &  7594781  &  19 08 35.91 &  +43 16 42.3 & 17.84(1) & 18.16(1) & 18.08(1) & 18.23(2) & 18.00(1) &  ZZ\,Ceti (a) \\ 
J1908+4619 &  9639485  &  19 08 25.69 &  +46 19 35.4 & 18.03(2) & 18.45(1) & 18.39(2) & 18.46(4) & 18.42(2) &  NOV $>$4.6\,ppt \\ 
J1909+4717 &  10198116 &  19 09 59.36 &  +47 17 09.5 & 15.82(1) & 16.30(1) & 16.25(1) & 16.43(1) & 16.28(1) &  NOV $>$0.13\,ppt (a,b) \\ 
J1911+4543 &  9272512  &  19 11 33.53 &  +45 43 46.1 & 18.26(2) & 18.67(2) & 18.65(2) & 18.73(5) & 18.69(3) &  - \\ 
J1913+4709 &  10132702 &  19 13 40.87 &  +47 09 30.6 & 18.71(3) & 19.08(3) & 19.01(3) & 19.15(8) & 18.93(4) &  ZZ\,Ceti (a) \\ 
J1917+3927 &  4357037  &  19 17 19.17 &  +39 27 18.2 & 17.91(1) & 18.28(1) & 18.22(1) & 18.38(3) & 18.17(2) &  ZZ\,Ceti (a) \\ 
J1917+4413 &  8293193  &  19 17 55.28 &  +44 13 26.2 & 18.10(2) & 18.42(1) & 18.34(2) & 18.47(4) & 18.27(3) &  ZZ\,Ceti \\
J1918+3914 &  -        &  19 18 54.33 &  +39 14 32.9 & 18.36(1) & 18.72(1) & 18.62(1) & 18.69(3) & 18.49(2) &  - \\ 
J1918+4533 &  9149300  &  19 18 30.15 &  +45 33 13.0 & 17.80(2) & 18.42(2) & 18.42(2) & 18.61(7) & 18.42(4) &  - \\ 
J1919+4247 &  -        &  19 19 55.09 &  +42 47 23.1 & 17.36(1) & 18.05(1) & 18.14(1) & 18.28(3) & 18.23(2) &  - \\ 
J1919+4712 &  10203164 &  19 19 52.92 &  +47 12 56.3 & 17.73(2) & 18.16(2) & 18.13(2) & 18.30(5) & 18.11(2) &  NOV $>$2.4\,ppt \\ 
J1919+4957 &  11759570 &  19 19 12.21 &  +49 57 51.3 & 17.40(1) & 17.92(1) & 17.96(2) & 18.18(6) & 17.93(3) &  - \\ 
J1920+4338 &  7885860  &  19 20 18.87 &  +43 38 32.4 & 18.52(3) & 18.84(2) & 18.71(2) & 18.93(5) & 18.66(3) &  - \\ 
J1920+5017 &  11911480 &  19 20 24.87 &  +50 17 22.1 & 17.68(2) & 18.13(2) & 18.05(2) & 18.13(6) & 18.05(4) &  ZZ\,Ceti (a,c)\\ 
J1922+4807 &  10794439 &  19 22 48.86 &  +48 07 21.4 & 18.05(2) & 18.70(2) & 18.74(3) & 18.84(7) & 18.71(4) &  - \\ 
J1923+3929 &  4362927  &  19 23 48.76 &  +39 29 33.1 & 19.05(2) & 19.43(1) & 19.33(2) & 19.44(6) & 19.26(3) &  ZZ\,Ceti \\ 
J1924+3655 &  1293071  &  19 24 08.27 &  +36 55 18.4 & 17.53(1) & 18.15(1) & 18.12(1) & 18.25(3) & 18.13(2) &  NOV $>$2.9\,ppt \\ 
J1926+3703 &  1432852  &  19 26 03.05 &  +37 03 16.4 & 18.49(1) & 19.04(1) & 19.04(2) & 19.24(5) & 19.03(3) &  NOV $>$6.9\,ppt \\ 
J1929+3857 &  3854110  &  19 29 52.07 &  +38 57 50.0 & 18.20(1) & 18.55(1) & 18.41(1) & 18.51(3) & 18.33(2) &  - \\ 
J1929+4302 &  -        &  19 29 12.27 &  +43 02 52.1 & 18.01(3) & 18.45(2) & 18.35(2) & 18.44(5) & 18.23(3) &  - \\ 
J1932+4210 &  6695659  &  19 32 12.04 &  +42 10 53.5 & 18.36(4) & 18.87(3) & 18.87(3) & 18.97(8) & 18.78(4) &  NOV $>$7.4\,ppt \\ 
J1933+4753 &  10604007 &  19 33 55.22 &  +47 53 02.4 & 18.71(3) & 19.13(2) & 19.03(2) & 19.13(7) & 18.92(4) &  - \\ 
J1935+4237 &  7124835  &  19 35 34.63 &  +42 37 41.7 & 17.31(1) & 17.76(1) & 17.70(1) & 17.83(3) & 17.69(2) &  - \\ 
J1935+4634 &  9775198  &  19 35 06.67 &  +46 34 59.1 & 17.88(1) & 18.28(1) & 18.20(1) & 18.36(3) & 18.15(2) &  NOV $>$7.4\,ppt \\  
J1939+4533 &  9162396  &  19 39 07.15 &  +45 33 33.9 & 18.15(2) & 18.51(1) & 18.48(2) & 18.65(5) & 18.36(2) &  ZZ\,Ceti \\ 
J1943+4538 &  -        &  19 43 02.67 &  +45 38 42.4 & 16.93(1) & 17.30(1) & 17.26(1) & 17.44(2) & 17.23(1) &  - \\ 
J1943+5011 &  -        &  19 43 31.92 &  +50 11 45.8 & 16.77(1) & 17.12(1) & 17.03(1) & 17.13(2) & 16.93(1) &  - \\ 
J1944+4327 &  7766212  &  19 44 05.85 &  +43 27 21.7 & 16.41(1) & 16.76(1) & 16.73(1) & 16.86(1) & 16.67(1) &  ZZ\,Ceti \\ 
J1945+4348 &  8043166  &  19 45 59.25 &  +43 48 45.3 & 17.26(1) & 17.61(1) & 17.55(1) & 17.58(2) & 17.45(1) &  - \\ 
J1945+4455 &  -        &  19 45 42.30 &  +44 55 10.6 & 16.82(1) & 17.16(1) & 17.09(1) & 17.20(2) & 16.99(1) &  ZZ\,Ceti \\ 
J1945+5051 &  12217892 &  19 45 03.58 &  +50 51 39.7 & 18.73(2) & 19.11(2) & 19.05(2) & 19.29(9) & 18.93(4) &  - \\ 
J1956+4447 &  -        &  19 56 16.98 &  +44 47 53.4 & 18.49(3) & 18.88(2) & 18.81(2) & 19.00(8) & 18.71(4) &  - \\ 
\hline
\end{tabular}
\\
(a) Observed by {\em Kepler} spacecraft; (b) \citet{Ostensen11b}; (c) \citet{Greiss14}
\end{table*}

\begin{table*}
\caption{Summary of spectroscopic observations and results from white dwarf model atmosphere fits to the spectra. \label{spec-summary}}
\scriptsize
\begin{tabular}{lrrllrrrr}
\hline
KIS ID     &  KIC ID   &  $g$ (mag)  &  Exp. time (s)     &  Night$^{\dagger}$     &  \teff\ (K)   & \logg\ (cgs) & Mass (\msun) & Comments \\
\hline
J1846+4157 &  -        &  18.3  &  1 $\times$ 1200  &  e    &  10\,770(130) & 8.03(0.05) & 0.62(03) &  - \\ 
J1848+4225 &  6923777  &  18.4  &  3 $\times$ 1200  &  e,h  &  12\,410(150) & 8.03(0.04) & 0.63(03) &  NOV $>$3.5\,ppt \\
J1851+4506 &  -        &  18.9  &  1 $\times$ 1500  &  e     &  9\,720(120) & 7.78(0.10) & 0.48(05) &  - \\ 
J1857+4908 &  11337510 &  18.8  &  1 $\times$ 1500  &  c    &  11\,160(160) & 8.27(0.07) & 0.77(04) & - \\ 
J1858+4613 &  9573820  &  18.8  &  1 $\times$ 1200  &  c    &  14\,800(1100) & 7.91(0.06) & 0.56(04) & - \\ 
J1859+4842 &  11125021 &  19.0  &  1 $\times$ 1200  &  c    &  11\,250(170) & 8.28(0.06) & 0.78(04) & NOV $>$6.3\,ppt \\ 
J1902+4223 &  -        &  19.0  &  1 $\times$ 1800  &  f    &  10\,640(130) & 8.05(0.05) & 0.63(03) & - \\ 
J1904+4130 &  -        &  17.5  &  1 $\times$ 900   &  e    &  11\,960(150) & 8.02(0.04) & 0.62(03) & - \\ 
J1904+4245 &  7184288  &  18.8  &  1 $\times$ 1200  &  c    &  13\,430(1440) & 8.54(0.06) & 0.95(04) & - \\ 
J1906+4354 &  8084967  &  19.0  &  1 $\times$ 1800  &  c    &  13\,890(520) & 7.94(0.05) & 0.58(03) & - \\ 
J1906+5002 &  11805054 &  19.3  &  1 $\times$ 1800  &  f    &  12\,980(240) & 8.01(0.06) & 0.61(04) & - \\ 
J1908+4316 &  7594781  &  18.2  &  6 $\times$ 1800  &  c,d,g & 11\,730(140)& 8.11(0.04) & 0.67(03) & ZZ\,Ceti (a) \\
J1908+4619 &  9639485  &  18.5  &  3 $\times$ 1200  &  c,h  &  14\,420(120)& 7.97(0.04) & 0.59(03) & NOV $>$4.6\,ppt \\  
J1909+4717 &  10198116 &  16.3  &  6 $\times$ 1500  &  a  &    13\,900(420) & 8.06(0.04) & 0.65(03) & NOV $>$0.13\,ppt (a,b) \\ 
J1911+4543 &  9272512  &  18.7  &  1 $\times$ 1200  &  c  &    13\,650(390) & 8.15(0.07) & 0.70(04) & - \\ 
J1913+4709 &  10132702 &  19.1  &  3 $\times$ 1500  &  d  &    11\,940(380) & 8.12(0.04) & 0.68(03) & ZZ\,Ceti (a) \\ 
J1917+3927 &  4357037  &  18.3  &  3 $\times$ 1200  &  d  &    10\,950(130) & 8.11(0.04) & 0.67(03) & ZZ\,Ceti (a) \\ 
J1917+4413 &  8293193  &  18.4  &  1 $\times$ 1200  &  c  &    12\,650(530) & 8.01(0.04) & 0.61(03) & ZZ\,Ceti \\
J1918+3914 &  -        &  18.7  &  1 $\times$ 1500  &  e  &    10\,960(140) & 7.95(0.05) & 0.57(03) & - \\ 
J1918+4533 &  9149300  &  18.4  &  1 $\times$ 1200  &  c  &    10\,570(130) & 7.98(0.06) & 0.59(03) & - \\ 
J1919+4247 &  -        &  18.1  &  1 $\times$ 1200  &  c  &    15\,440(490) & 8.36(0.06) & 0.84(04) & - \\ 
J1919+4712 &  10203164 &  18.1  &  1 $\times$ 900   &  c  &    10\,400(230) & 8.27(0.07) & 0.77(04) & NOV $>$2.4\,ppt \\ 
J1919+4957 &  11759570 &  18.0  &  5 $\times$ 1275  &  b  &    14\,470(260) & 8.22(0.13) & 0.75(06) & - \\ 
J1920+4338 &  7885860  &  18.7  &  1 $\times$ 1800  &  c  &    10\,330(130) & 7.95(0.06) & 0.57(04) & - \\ 
J1920+5017 &  11911480 &  18.1  &  3 $\times$ 1200  &  e  &    11\,580(140) & 7.96(0.04) & 0.58(03) & ZZ\,Ceti (a,c)\\ 
J1922+4807 &  10794439 &  18.7  &  1 $\times$ 1200  &  c  &    10\,140(120) & 8.12(0.08) & 0.67(04) & - \\ 
J1923+3929 &  4362927  &  19.4  &  1 $\times$ 1800  &  f  &    11\,140(140) & 7.84(0.05) & 0.51(03) & ZZ\,Ceti \\ 
J1924+3655 &  1293071  &  18.1  &  5 $\times$ 1200  &  b,c,g & 11\,550(140)& 7.99(0.04) & 0.60(03) & NOV $>$2.9\,ppt \\
J1926+3703 &  1432852  &  19.0  &  1 $\times$ 1200  &  c  &    12\,240(360) & 8.82(0.07) & 1.11(04) & NOV $>$6.9\,ppt \\ 
J1929+3857 &  3854110  &  18.4  &  1 $\times$ 1200  &  c  &    10\,480(130) & 7.85(0.09) & 0.52(05) & - \\ 
J1929+4302 &  -        &  18.4  &  2 $\times$ 1200  &  c  &    10\,500(130) & 7.86(0.08) & 0.52(04) & - \\ 
J1932+4210 &  6695659  &  18.9  &  1 $\times$ 1500  &  f  &    11\,560(140) & 7.83(0.06) & 0.51(04) & NOV $>$7.4\,ppt \\ 
J1933+4753 &  10604007 &  19.0  &  1 $\times$ 1500  &  e  &    10\,140(120) & 8.31(0.07) & 0.80(04) & - \\ 
J1935+4237 &  7124835  &  17.7  &  1 $\times$ 900   &  f  &    13\,560(420) & 7.86(0.04) & 0.53(03) & - \\ 
J1935+4634 &  9775198  &  18.2  &  1 $\times$ 1200  &  f  &    12\,930(200) & 8.06(0.05) & 0.64(03) & NOV $>$7.4\,ppt \\  
J1939+4533 &  9162396  &  18.5  &  1 $\times$ 1200  &  e  &    11\,070(140) & 8.06(0.05) & 0.64(03) & ZZ\,Ceti \\ 
J1943+4538 &  -        &  17.3  &  1 $\times$ 900   &  f  &    13\,120(380) & 7.90(0.04) & 0.55(03) & - \\ 
J1943+5011 &  -        &  17.0  &  1 $\times$ 900   &  d  &    10\,730(130) & 8.08(0.04) & 0.65(03) & - \\ 
J1944+4327 &  7766212  &  16.7  &  3 $\times$ 900   &  d,g  &  11\,890(150) & 8.01(0.04) & 0.61(03) & ZZ\,Ceti \\ 
J1945+4348 &  8043166  &  17.6  &  1 $\times$ 900   &  f  &    10\,430(130) & 8.05(0.04) & 0.63(03) & - \\ 
J1945+4455 &  -        &  17.1  &  1 $\times$ 900   &  e  &    11\,590(140) & 8.04(0.04) & 0.63(03) & ZZ\,Ceti \\ 
J1945+5051 &  12217892 &  19.1  &  1 $\times$ 1800  &  e  &    11\,910(380) & 7.98(0.06) & 0.59(04) & - \\ 
J1956+4447 &  -        &  18.8  &  1 $\times$ 1500  &  e  &    14\,160(820) & 7.91(0.05) & 0.56(03) & - \\ 
\hline
\end{tabular}
\\
$^{\dagger}$ See Table~\ref{tab:specjour}; (a) Observed by {\em Kepler} spacecraft; (b) \citet{Ostensen11b}; (c) \citet{Greiss14}
\end{table*}

\begin{figure}
\includegraphics[width=0.98\columnwidth]{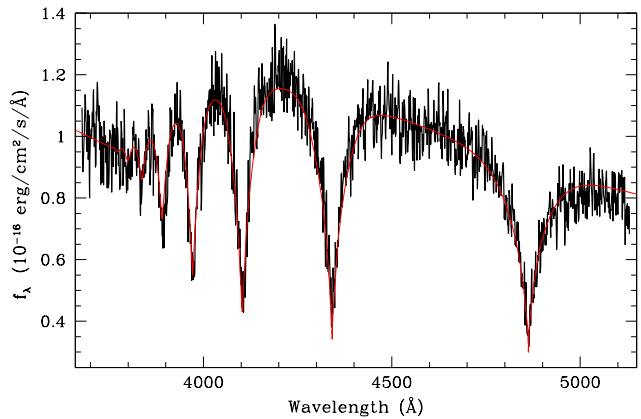}
\caption{\small A representative WHT/ISIS spectrum from our sample, KISJ1923+3929 ($g=19.4$\,mag, S/N$\simeq$12), obtained in a single 1800\,s exposure on 2013 June 8 confirming the star to be a DA white dwarf. The best-fit atmosphere parameters (\teff\ $=11\,160\pm140$\,K, \logg\ $=7.90\pm0.04$), shown in red and detailed in Table~\ref{spec-summary}, place the star within the empirical ZZ\,Ceti instability strip, and pulsations were subsequently detected from the ground (see Figure~\ref{fig:lcnewzzceti}). \label{spec-fit}}
\end{figure}

We fitted one-dimensional model atmospheres to our spectra in order to obtain their effective temperatures and surface gravities. To derive the most robust atmospheric parameters, we have fitted the six Balmer lines H$\beta-$H9 using synthetic spectra computed by two independent groups. The first fits use the pure hydrogen atmosphere models and fitting procedure described in \citet{Tremblay11b} and references therein, which employ the ML2/$\alpha = 0.8$ prescription of the mixing-length theory \citep{Tremblay10}. The second set of fits use the pure hydrogen atmosphere models detailed in \citet{Koester10}, which also employ ML2/$\alpha = 0.8$, following the procedure described in \cite{Napiwotzki04}. Since the empirical instability strip has most thoroughly been defined by the first set of models, our final determinations use the models described in \citet{Tremblay11b}. If the parameters found from the \citet{Koester10} models differ by more than 1-$\sigma$, we have increased the parameter uncertainties to compensate for this disagreement.

When multiple exposures were taken for a given star, we calculated the atmospheric parameters for each individual spectrum, and then took the weighted mean as a final value. The results from the fits to the spectra of all our observed white dwarfs are given in Table~\ref{spec-summary}. These parameters have been corrected for the three-dimensional dependence of convection described by \citet{Tremblay13}. Additionally, we include in Table~\ref{spec-summary} estimated spectroscopic masses of our 43 white dwarfs, found using the mass-radius relation and the evolutionary cooling models from \cite{Fontaine01} with a carbon-oxygen core \cite{Bergeron01}\footnote{The cooling models can be found on http://www.astro.umontreal.ca/$\sim$bergeron/CoolingModels/. Also refer to \cite{HB06, Tremblay11b, Bergeron11} for colour and model calculations.}.

\begin{figure*}
\centering{\includegraphics[width=0.97\textwidth]{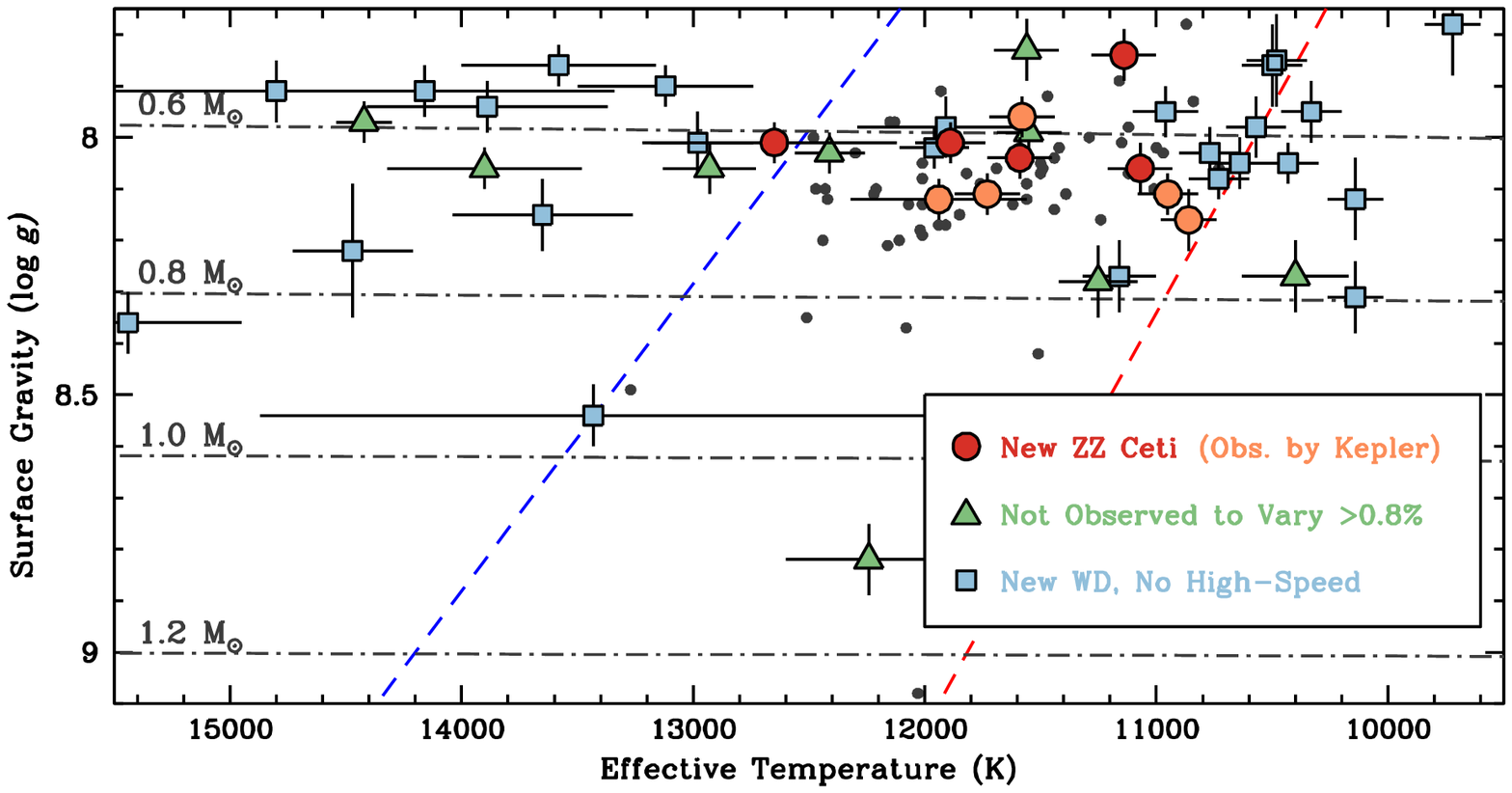}}
\caption{($T_\mathrm{eff}$, $\log g$) diagram of the white dwarfs we observed spectroscopically with WHT/ISIS. The red circles correspond to new ZZ\,Ceti stars; the five orange circles are ZZ\,Cetis ultimately observed by {\em Kepler} and include one not characterized here, KIC\,4552982 \citep{Bell15}. The objects marked as green triangles were observed not to vary to at least less than 0.8\,ppt amplitude, with limits detailed in Table~\ref{astero-summary}. The objects marked as light blue squares have not been observed with high-speed photometry. The small, dark grey points correspond to known ZZ\,Ceti stars with atmospheric parameters characterized in an identical way \citep{Gianninas11,Tremblay11b}, and the dark grey dashed-dotted lines show cooling tracks of various masses \citep{Fontaine01}. The red and blue dashed lines correspond to the latest boundaries of the empirical ZZ\,Ceti instability strip \citep{Tremblay15}. All parameters have been corrected for the 3D dependence of convection \citep{Tremblay13}. \label{kiel-diag}}
\end{figure*}

In Figure\,\ref{spec-fit}, we show the WHT spectrum of one of our nine new confirmed ZZ\,Ceti stars, overplotted with the best-fit atmospheric parameters. Several of the most promising candidates were followed up with ground-based time-series photometry, which we describe in the following section. Figure~\ref{kiel-diag} shows our atmospheric parameters and subsequent high-speed photometry in the context of the ZZ\,Ceti instability strip.

\section{Ground-based high-speed photometry}
\label{astero}

We obtained ground-based optical time-series photometry for 17 of our 43 ZZ\,Ceti candidates, in order to check for variability. The ground-based observations of eight of our stars were obtained using the Wide Field Camera (WFC) mounted on the 2.5-m Isaac Newton Telescope on La Palma. These INT observations in 2012 were part of the RATS-{\it Kepler} survey \citep{Ramsay14}, which is a deep optical high-cadence survey of objects in the {\it Kepler} field using one-hour-long sequences of 20\,s $g$-band exposures.

Additionally, we obtained high-speed photometry using the Argos \citep{Nather04} and later the ProEM camera mounted on the 2.1-m Otto Struve Telescope at McDonald Observatory. These McDonald observations were obtained through a 3\,mm {\em BG40} filter to reduce the effects of transparency variations, and were reduced using the external {\sc iraf} package $\textit{ccd\_hsp}$ written by Antonio Kanaan \citep{Kanaan02}. For every light curve we divided the normalized, sky-subtracted target flux by the measured flux for the brightest comparison star of similar colour in the field.

In order to assess variability in these white dwarf stars, we computed a relatively conservative significance threshold following the method outlined in \citet{Greiss14}. In short, we randomly permuted the fluxes for each time in the light curve, performed a Fourier transform (FT), recorded the highest peak in the resultant FT, and repeated the process $10\,000$ times. We derived a 3$\sigma$ threshold as the value for which 9970 of these permuted FTs had a lower maximum amplitude. We considered any peaks in the FT in the original light curve above this value significant. The 3$\sigma$ threshold also marks a useful limit for the maximum pulsation amplitudes we would be sensitive to for the white dwarfs which we designate Not Observed to Vary (NOV).

We include a full summary of this high-speed photometry and the appropriate 3$\sigma$ limits in Table~\ref{astero-summary}. Our relative amplitude units are in parts per thousands (ppt), where 1\,ppt = 0.1\,per\,cent. 

\subsection{Confirmation of Nine New ZZ\,Cetis}
\label{sec:confirmationnewzzcetis}

\begin{figure}
\centering{\includegraphics[width=0.97\columnwidth]{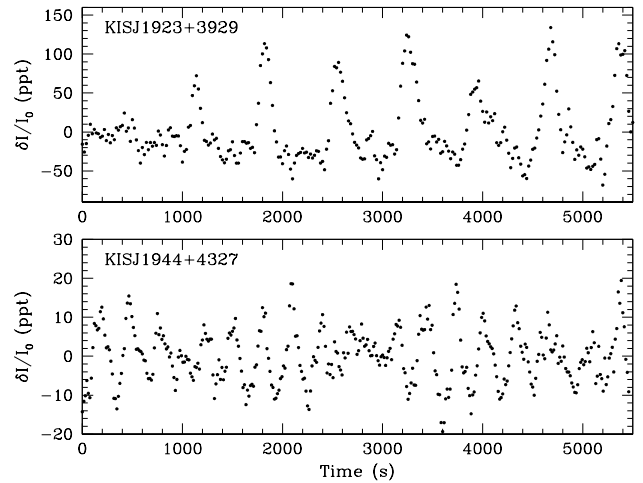}}
\caption{Light curves of two of our new ZZ\,Cetis observed from McDonald Observatory. The top panel shows a relatively cool ZZ\,Ceti, KISJ1923+3929, with a highly non-sinusoidal pulse shape and dominant oscillation period of roughly 724\,s. The bottom panel shows the brightest new ZZ\,Ceti we confirmed, KISJ1944+4327, with much lower-amplitude and shorter pulsation periods of 321.95 and 274.76\,s. 1\,ppt = 0.1\,per\,cent. \label{fig:lcnewzzceti}}
\end{figure}

Nine of the 17 white dwarfs we have followed up with high-speed photometry show significant photometric variations at periods consistent with $g$-mode pulsations in typical white dwarfs \citep{Mukadam04}. Some pulsational variability was immediately evident in the light curve -- Figure~\ref{fig:lcnewzzceti} shows two of our new ZZ\,Ceti stars observed from McDonald Observatory.

\begin{figure}
\centering{\includegraphics[width=0.97\columnwidth]{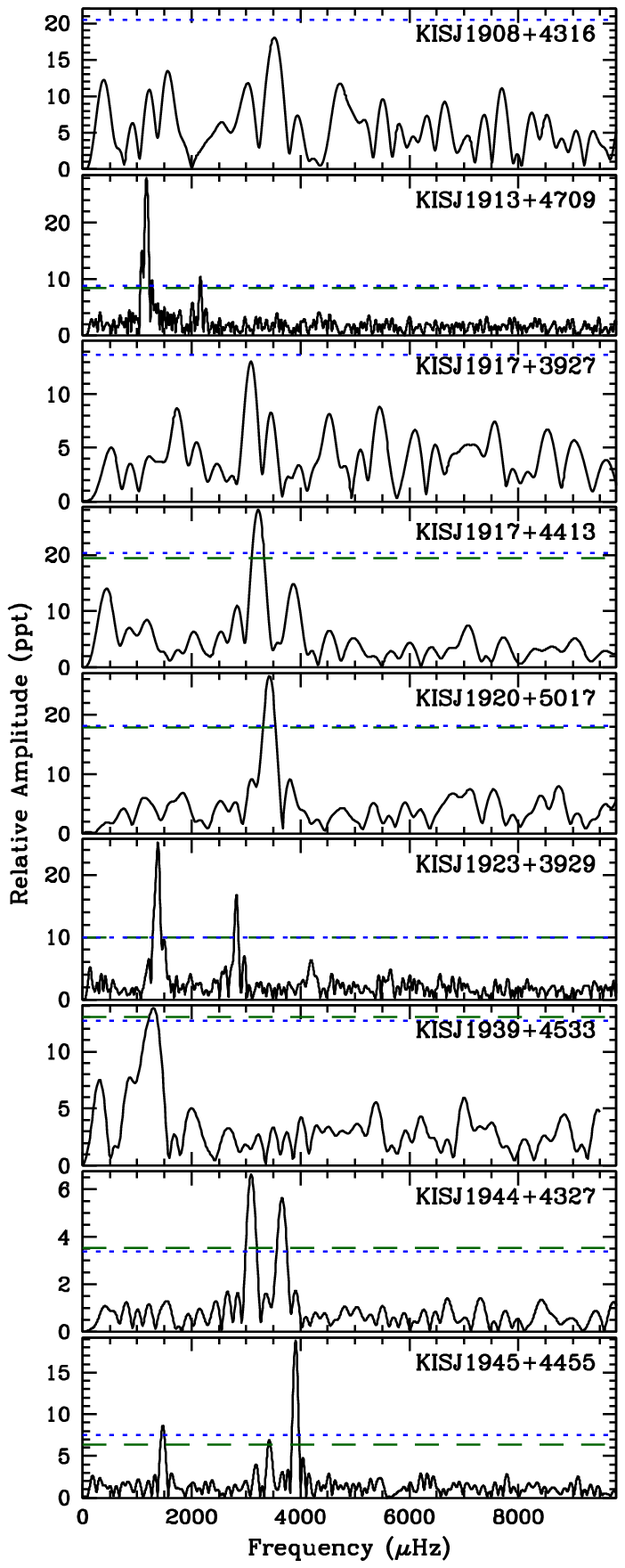}}
\caption{Fourier transforms confirming variability in seven of our nine new ZZ\,Ceti stars. The dotted blue line corresponds to the 4$\langle {\rm A}\rangle$ line marking four times the mean FT amplitude. The dashed green line shows the 3$\sigma$ significance line, as described in the text and listed in Table~\ref{astero-summary}. The targets KISJ1908+4316 and KISJ1917+3927 showed promising light curves, and were followed up and confirmed as ZZ\,Ceti stars with {\em Kepler} space-based photometry; their analysis is presented in a forthcoming paper. \label{fig:ftnewzzceti}}
\end{figure}

Figure~\ref{fig:ftnewzzceti} shows FTs of all nine new ZZ\,Cetis, and the top half of Table~\ref{astero-summary} summarizes the characteristics of the highest-amplitude variability in each white dwarf. Two white dwarfs observed as part of the RATS survey did not show significant peaks (KISJ1908+4316 and KISJ1917+3927), but were later confirmed to pulsate using extended {\em Kepler} observations.

As white dwarfs cool through the ZZ\,Ceti instability strip, their convection zones deepen, driving longer-period pulsations \citep{Winget82, Tassoul90}. This trend is borne out in our time-series photometry. Most of the new ZZ\,Ceti stars found by our work have dominant pulsations between $250-320$\,s, typical of objects within the ZZ\,Ceti instability strip. Of note, the three new ZZ\,Ceti stars with main pulsation periods $>720$\,s all have spectroscopically determined effective temperatures $<11\,550$\,K, putting them near the cooler, red edge of the instability strip.

Once we were able to establish pulsations from ground-based follow-up, we submitted the target for space-based, minute-cadence monitoring by the {\em Kepler} spacecraft. Since all ZZ\,Cetis we discovered are relatively faint (all have $g>16.7$\,mag, see Table~\ref{spec-summary}), extended time-series photometry will significantly improve the signal-to-noise of the observed pulsation spectrum and enable the detection of weak pulsation modes, which is not easily feasible through ground-based studies, given aliasing.

However, only four of our targets were observed from space, and two for just one month, before {\em Kepler} lost its ability to maintain fine pointing at its original field near Cygnus due to the failure of the second and critical reaction wheel. The four with space-based data obtained thanks to this program are KISJ1920+5017 (KIC\,11911480, \citealt{Greiss14}), KISJ1908+4316 (KIC\,07594781), KISJ1917+3927 (KIC\,04357037), and KISJ1913+4709 (KIC\,10132702). We will present analysis of the latter three ZZ\,Ceti stars in a forthcoming publication.

\renewcommand{\arraystretch}{1.2}
\begin{table*}
\caption{Summary of ground-based, time-series photometry of 17 ZZ\,Ceti candidates from our survey. We include the dominant periods and their amplitudes of our nine new confirmed ZZ\,Cetis. Targets marked with a $^{\dagger}$ were followed up by the {\em Kepler} space telescope. \label{astero-summary}}
\scriptsize
\begin{tabular}{lrrrrrrrrr}
\hline \hline
KIS ID     & KIC ID   &  $g$ (mag) & Period (s) & Amp. (ppt) & 3$\sigma$ (ppt) &  Telescope, Obs. Date & Filter & Length (hr) & Exp. (s) \\
\hline
J1908+4316 & 7594781  &  18.2 & 283.8(3.0) & 17.8(4.4)$^{\dagger}$ & 20.5  & 2.5m\,INT, 2012~Aug~05 & SDSS-$g$ &  1.0 & 49 \\
J1913+4709 & 10132702 &  19.1 & 853.5(1.1) & 28.1(1.4)$^{\dagger}$ & 8.9   & 2.1m\,McD, 2012~Jun~20 & BG40 &  5.1 & 20 \\
J1917+3927 & 4357037  &  18.3 & 323.4(3.5) & 13.0(2.9)$^{\dagger}$ & 13.7  & 2.5m\,INT, 2012~Aug~10 & SDSS-$g$ &  1.0 & 48 \\
J1917+4413 & 8293193  &  18.4 & 310.9(1.5) & 27.9(3.1)             & 20.4  & 2.5m\,INT, 2012~Aug~11 & SDSS-$g$ &  1.0 & 48 \\
J1920+5017 & 11911480 &  18.1 & 291.5(1.3) & 26.5(2.9)$^{\dagger}$ & 18.1  & 2.5m\,INT, 2011~Aug~01 & SDSS-$g$ &  1.1 & 49 \\
J1923+3929 & 4362927  &  19.4 & 723.6(1.1) & 25.3(1.7)             & 10.0  & 2.1m\,McD, 2012~Jun~21 & BG40 &  4.6 & 20 \\
J1939+4533 & 9162396  &  18.5 & 766(12)    & 14.1(2.5)             & 12.7  & 2.5m\,INT, 2014~Jun~18 & SDSS-$g$ & 1.3 & 51 \\
J1944+4327 & 7766212  &  16.7 & 321.95(84) & 6.71(54)              & 3.4  & 2.1m\,McD, 2014~Jul~25 & BG40 & 1.5 & 15 \\
J1945+4455 &  -       &  17.1 & 255.92(17) & 19.0(1.0)             & 7.5   & 2.5m\,INT, 2014~Jun~17 & SDSS-$g$ & 2.9 & 35 \\
\hline
J1848+4225 & 6923777  &  18.4 & -         & NOV       & 3.5     & 2.1m\,McD, 2014~Aug~04 & BG40 & 4.0 & 20 \\
J1859+4842 & 11125021 &  19.0 & -         & NOV       & 6.3     & 2.1m\,McD, 2014~Sep~03 & BG40 & 5.0 & 20 \\
J1908+4619 & 9639485  &  18.5 & -         & NOV       & 4.6     & 2.1m\,McD, 2014~Sep~04 & BG40 & 3.0 & 15 \\
J1919+4712 & 10203164 &  18.1 & -         & NOV       & 2.4     & 2.1m\,McD, 2014~Jul~22 & BG40 & 8.2 & 5 \\
J1924+3655 & 1293071  &  18.1 & -         & NOV       & 2.9     & 2.1m\,McD, 2014~Jul~25 & BG40 & 3.9 & 10 \\
J1926+3703 & 1432852  &  19.0 & -         & NOV       & 6.9     & 2.1m\,McD, 2015~Oct~12 & BG40 & 3.5 & 10 \\
J1932+4210 &  6695659 &  18.9 & -         & NOV       & 6.1     & 2.1m\,McD, 2015~Jun~11 & BG40 & 2.8 & 10 \\
           &          &       & -         & NOV       & 7.4     & 2.5m\,INT, 2014~Jun~20 & SDSS-$g$ & 1.7 & 67 \\
J1935+4634 &  9775198 &  18.2 & -         & NOV       & 7.4     & 2.5m\,INT, 2014~Jun~20 & SDSS-$g$ & 1.5 & 65 \\
\hline
\end{tabular}
\end{table*}
\renewcommand{\arraystretch}{1}

\subsection{White dwarfs not observed to vary}

\begin{figure}
\centering{\includegraphics[width=0.97\columnwidth]{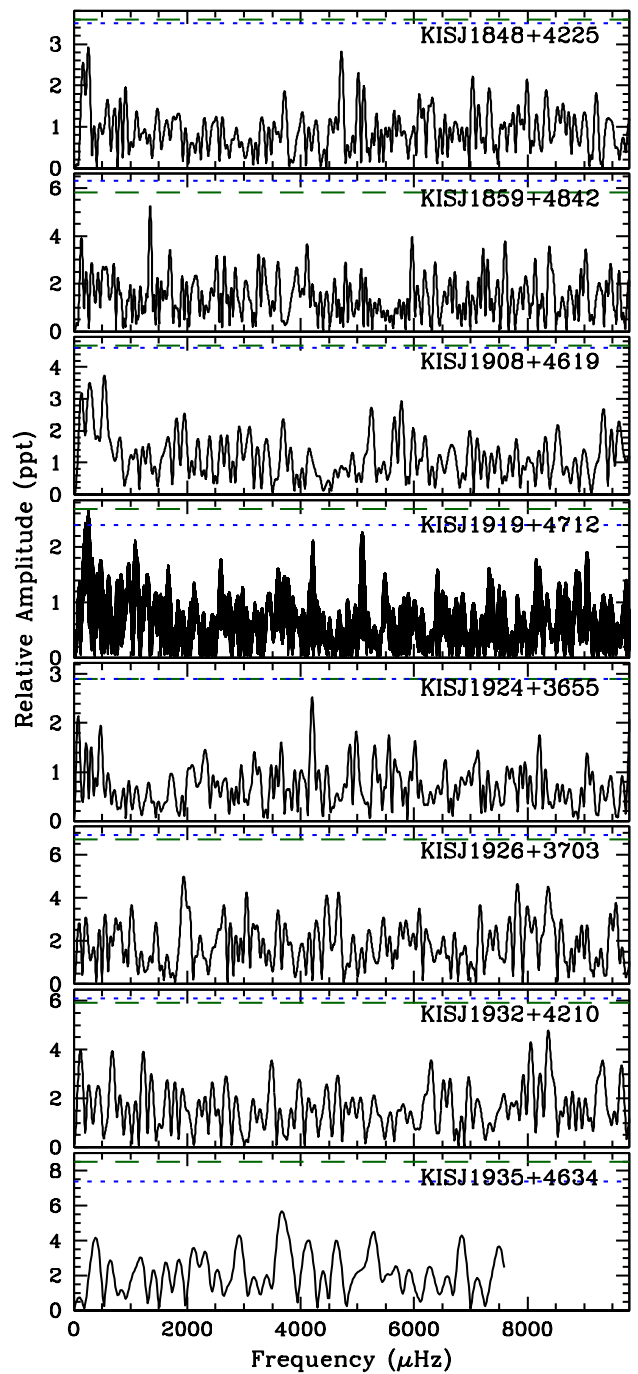}}
\caption{Fourier transforms of eight pulsating white dwarf candidates observed not to vary. The blue dotted line and green dashed lines have the same meaning as before.  \label{fig:ftnewnov}}
\end{figure}

In addition to the confirmation of nine new ZZ\,Ceti stars, our ground-based photometry has put relatively strong limits on the lack of photometric variability in eight other white dwarfs with effective temperatures near the ZZ\,Ceti instability strip. Those limits are quoted as a 3$\sigma$ threshold in Table~\ref{astero-summary}, and represented visually with a FT for each object in Figure~\ref{fig:ftnewnov}.

Several of these non-variable white dwarfs have atmospheric parameters that place them within the empirical ZZ\,Ceti instability strip given their uncertainties, most recently updated by \citet{Tremblay15}. It has been assumed that all white dwarfs in this region can foster a partial ionization zone and thus pulsate --- that the instability strip is pure --- but this claim is still under review \citep{Bergeron04,Mukadam04b,Castanheira07}, although there is good agreement between the observed and theoretical ZZ Ceti instability strip (e.g., \citealt{VanGrootel13}).

We will not address that issue here, other than to say there are a number of reasons why these objects may appear not to pulsate. Firstly, it is possible the white dwarfs really do vary, but at lower amplitudes than our detection limits allow (e.g., \citealt{Castanheira10}). Additionally, it is possible that subtle issues in the flux calibration of our spectra have introduced unaccounted for systematic uncertainties in the derived atmospheric parameters, and the stars may have true temperatures outside the instability strip. Given that the non-variable stars are more or less uniformly distributed in the ($T_\mathrm{eff}$, $\log g$) plane, it is likely that the low signal-to-noise of the time-averaged spectra is responsible for some interlopers, as demonstrated quantitatively by \citet{Gianninas05}.

The most interesting white dwarf not observed to vary in our sample is the ultramassive J1926+3703, which sits in the middle of the empirical instability strip. At $1.14\pm0.04$\,\msun, it would be the second-most-massive white dwarf known to pulsate, behind only GD\,518 \citep{Hermes13}. White dwarfs this massive likely have at least partially crystallized interiors. We did not detect photometric variability in this relatively faint target ($g=19.0$\,mag) to a limit of roughly 6.9\,ppt, but it is quite possible that this white dwarf pulsates at lower amplitude. For example, the two most massive known pulsating white dwarfs have low pulsations amplitudes, which rarely exceed 5\,ppt in a given night and are often far lower in amplitude \citep{Kanaan05,Hermes13}.

The {\em Kepler} mission would have given at least an order-of-magnitude improvement on these NOV limits, were these targets observed from space. For example, \citet{Ostensen11b} showed that J1909+4717 did not vary to at least an amplitude of 0.13\,ppt using just one month of {\em Kepler} data in Q4.1. However, only the four known pulsating white dwarfs listed at the end of Section~\ref{sec:confirmationnewzzcetis} were observed before the spacecraft had a failure of its second reaction wheel.

We note that many white dwarfs with atmospheric parameters within the empirical instability strip remain unobserved, and we encourage additional follow-up to constrain pulsational variability in these white dwarfs. We especially encourage further monitoring of the ultramassive J1926+3703.

\section{Conclusion}
\label{conclusion}

Using photometric colour selection from the KIS survey and subsequent medium-resolution spectroscopy from ISIS on the WHT, we have discovered and characterized 42 new white dwarfs in the original {\it Kepler} mission field. Follow-up, high-speed photometry from ground-based telescopes confirmed that at least nine of these objects are ZZ\,Ceti stars. Four were subsequently observed from space using the {\em Kepler} space telescope, and \citet{Greiss14} report on the first ZZ\,Ceti found in this project, KIC\,11911480. Asteroseismic inferences from the extended datasets on the other three will be presented in a forthcoming publication. 

Unfortunately, the failure of the second and critical reaction wheel that kept {\em Kepler} precisely pointed towards its original mission field occurred within months of our discovery of many of these new ZZ\,Ceti stars, and our sparse ground-based discovery light curves are so far the only time-series photometry for most of the targets here. At the end of its roughly four years pointed towards its original field, {\em Kepler} observed a total of six pulsating white dwarfs.

However, the two-reaction-wheel controlled {\em K2} mission will cover a significantly larger footprint as it tours the ecliptic \citep{Howell14}, and {\em K2} will likely observe more than a thousand white dwarfs, dozens of them pulsating. Already several important discoveries have come from these {\em K2} observations of pulsating white dwarfs \citep{Hermes14,Hermes15a,Hermes15b}. The colour selection methods honed in this work have ensured that dozens of pulsating white dwarfs have been or will be observed by {\em K2} for up to 80\,d at a time, using selection from a variety of multiwavelength photometric surveys.

\section*{Acknowledgments}

The authors acknowledge fruitful conversations with A. Gianninas during the preparation of this manuscript. The research leading to these results has received funding from the European Research Council under the European Union's Seventh Framework Programme (FP/2007-2013) / ERC Grant Agreement n. 320964 (WDTracer). Support for this work was provided by NASA through Hubble Fellowship grant \#HST-HF2-51357.001-A, awarded by the Space Telescope Science Institute, which is operated by the Association of Universities for Research in Astronomy, Incorporated, under NASA contract NAS5-26555. D.S. acknowledges support from STFC through an Advanced Fellowship (PP/D005914/1) as well as grant ST/I001719/1. K.J.B. acknowledges funding from the NSF under grants AST-0909107 and AST-1312983, the Norman Hackerman Advanced Research Program under grant 003658-0252-2009, and the {\em Kepler} Cycle 4 Guest Observer program 11-KEPLER11-0050. This paper makes use of data collected at the Isaac Newton Telescope and the William Herschel Telescope, both operated on the island of La Palma, by the Isaac Newton Group in the Spanish Observatorio del Roque de los Muchachos, as well as data taken at The McDonald Observatory of The University of Texas at Austin.

%%%%%%%%%%%%%%%%%%%%%%%%%%%%%%%%%%%%%%%%%%%%%%%%%%

% Don't change these lines
\bsp	% typesetting comment
\label{lastpage}
\end{document}